\RequirePackage{ifpdf}

\documentclass{pCHEF}
\usepackage{subcaption}
\usepackage{url}

\title{First results of the SDHCAL technological prototype}

\ShortTitle{SDHCAL first results}

\author{\speaker{Y. Haddad}, On behalf of the CALICE Collaboration\\
  \'Ecole polytechnique, LLR ,CNRS-IN2P3\\
  E-mail: \email{yacine.haddad@cern.ch}}
\abstract{ 
  The CALICE Semi-digital hadronic calorimeter built in 2011, was installed and tested during two periods of two weeks each in 2012 at CERN SPS facilities. The detector has more than 450000 channels with a semi-digital readout distributed on 48 layers with efficiency exceeding $95\%$. It has been run using the trigger-less and power pulsing modes. Data have been collected with muon, electron and hadron beams in the energy range between $5$ and $80~\GeV$. This contribution focuses on the performances, the shower selection methods and on the first results on the calibration using pions.
}

\FullConference{Calorimetry for High Energy Frontiers - CHEF 2013,\\
  April 22-25, 2013\\
  Paris, France}

\usepackage{CHEF2013_definitions}

\begin{document}
\section*{Introduction}\label{sec:intro}

In order to investigate physics phenomena at energy ranging up to the TeV scale, the future international $e^+e^-$ linear collider, (ILC) \cite{ILCTDR:Detector} will need to perform precise measurements of momentum and energy of jets, in events where Standard Model bosons are produced in association with hadronic final states. The method proposed for ILC detectors to address this goal is so called ``\textit{Particle Flow Algorithm}'' (PFA)\cite{brientPFA}\cite{Videau:2002sk} approach. It consists in tracking the response of each particle, produced in the collision, in every sub-detector, and combining them to reach the highest precision. This method requires electromagnetic and hadronic calorimeter (ECAL and HCAL respectively) with tracking capability in addition to their standard energy measurement.

A gaseous digital hadron calorimeter (DHCAL) \cite{Ammosov:2002jq} can provide a fine segmentation ($1~cm^2$). It is based on $1-\rm bit$ readout system, and hit counting for the energy estimation. However, at high energy, this approach suffers from saturation effect. A $2-bit$ readout (semi-digital readout) allows for three thresholds \cite{SDHCALDev}, can correct this effect and improve the energy resolution at high energy. This option is called semi-digital HCAL (SDHCAL).

%For the SDHCAL prototype \cite{CHEF:Grenier}, the energy measurement can be achieved in the first order by a simple counting of cells fired by the incoming particle. At the second order, weighted means between 3 thresholds can improve the precision.

In this paper the results of exposure of the SDHCAL prototype to pion beam in the energy range $5-80\GeV$ at CERN H2 and H6 beam lines\cite{CERN:beamLine} are reported. The experimental set-up as well as the event reconstruction and data quality control are first briefly described. Then, the methods for particle identification and event selection are highlighted. Finally, results on SDHCAL response to single particle are presented and commented upon.

\section{The $\mathbf{1~m^3}$ SDHCAL Technological prototype \& 
  CERN beam test}\label{sec:sdhcal}

A technological prototype was build to validate the SDHCAL\cite{CHEF:Grenier} concept. It is composed by 48 active layers and 51 stainless-steel absorbers (equivalent to $6~\lambda_I$). Each sensitive layer is made of a $1\times1~\rm m^2$ Glass Resistive Plate Chamber (GRPC) \cite{GRPCPerf}, segmented in $1\times 1 \cm^2$ readout pads, for grand total over $450000$ channels for the full detector. The chambers are filled with a gas mixture (TFE: $93\%$, $\rm CO_{2}$: $5\%$, $\rm SF_6$ : $2\%$) polarized by a difference of potential of about $7~\rm kV$. The thresholds were set at $0.114~\rm$, $5~\rm$ and $15~\rm pC$ corresponding to 0.1, 4 and 12.5 mip \footnote{``mip'' is the mean deposit energy of the minimum ionizing particle. In the case of the GRPC chambers $1~\rm mip$ correspond to an induced charge of $1.1~\rm pC$} respectively. No gain correction was applied.

Several test beam campaigns were engaged. The results presented in this paper are based on the data saved during August and September 2012 near the SPS at CERN. The SPS beam line facilities provide a beam with an energy knows with $\delta E_{beam}/E_{beam}\sim 1\%$ of precision. The low acceptance of the GRPC detectors at high particle rate (must be $\leq 100\rm Hz/cm^2$), requires enlarging the beam profile.

\section{Reconstruction \& data quality control}
\subsection{Event Building}\label{sec:trivent}

The so called trigger-less mode was used for the data acquisition;  In this mode each very front end chip (treating 64 channels) auto-triggers and stores the information. The acquisition system reads the full detector when the memory of at least one memory is full. The collected data thus include not only the information about the incoming particles (pion, muon, cosmic ... ) but also the noise. The average duration of one acquisition window is of $\sim 30~\rm ms$. Within each window, the time of each hit w.r.t the start of the acquisition is recorded by a counter (``time-stamp'') increasing by steps of $200~\rm ns$.

\begin{figure}[!htbp]
  \centering
  \begin{subfigure}[b]{0.48\textwidth}
    \centering
    \includegraphics[width=8cm,keepaspectratio]{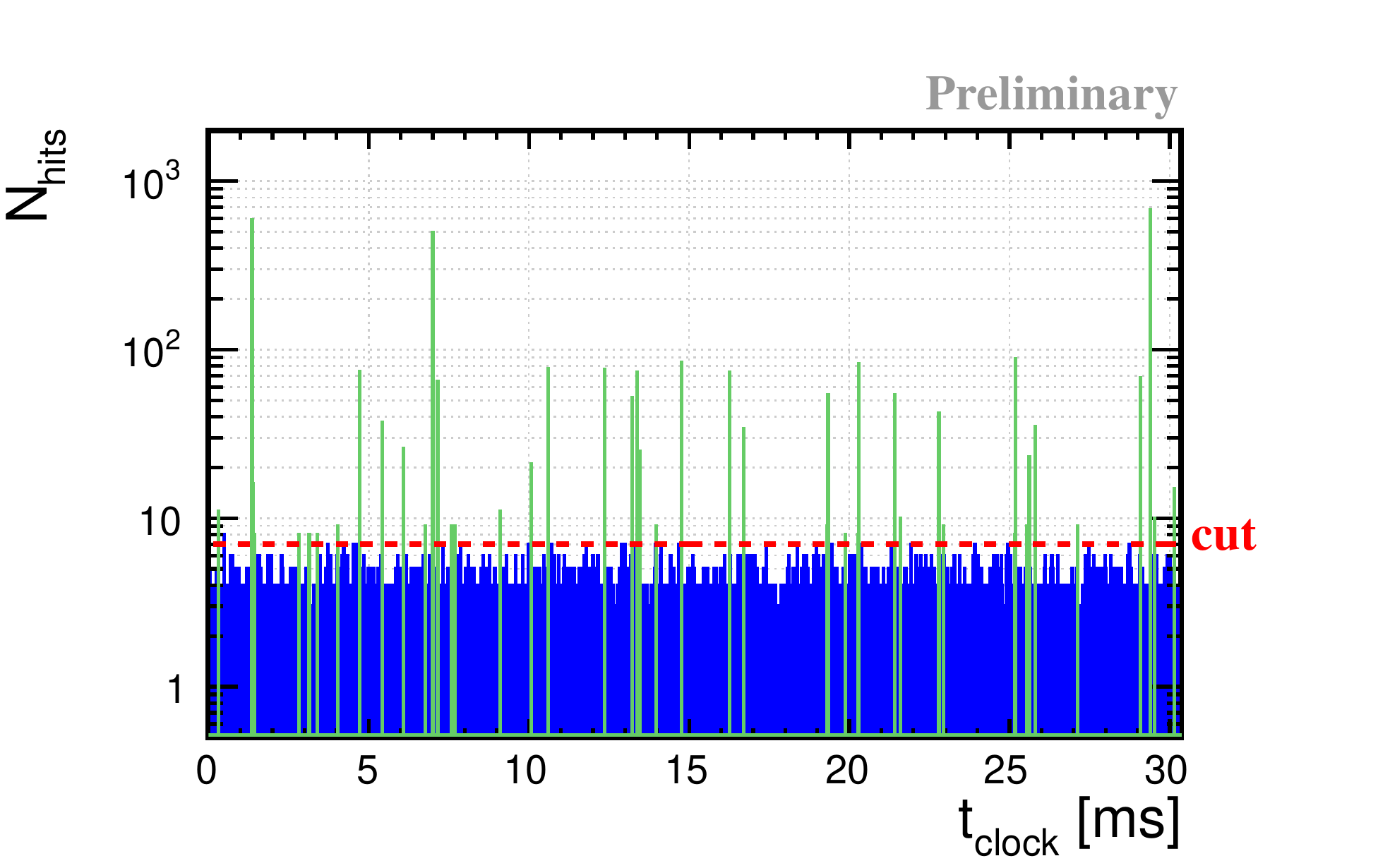}
    \caption{ }
    \label{fig:time_spectrum}
  \end{subfigure}%
  \hspace{1cm}
  \begin{subfigure}[b]{0.38\textwidth}
    \centering 
    \includegraphics[width=7cm,keepaspectratio]{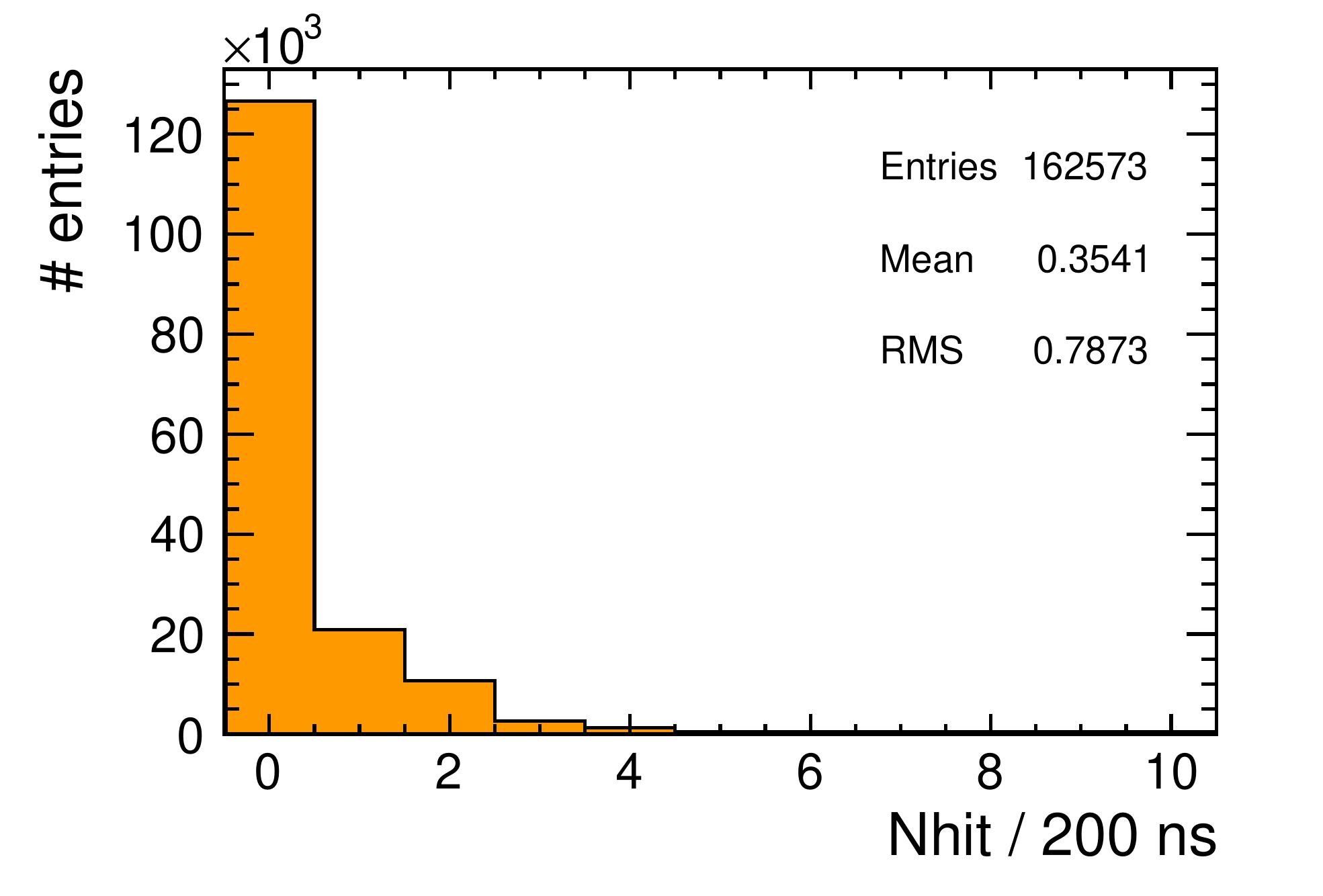}
    \caption{ }
    \label{fig:noise_out_event}
  \end{subfigure}
  \caption{ ({\bf a}) Time spectrum of an acquisition window with a granularity of $200~\rm ns$. The candidate physical events are highlighted over the background noise (in blue). ({\bf b}) Distribution per clock tick of number of intrinsic noisy hits.}
\end{figure}

The physical events are built from hit collection using a time clustering method. A histogram of hits time occurrence is filled for each acquisition readout (figure \ref{fig:time_spectrum}) with a bin-width set to the time-stamp precision. Only clock ticks with a number of hits higher than $7$ are used to initiate the time clustering process. Hits belonging to the adjacent clock ticks are combined to build a physical event. Care was taken to ensure that no hit belongs to two different events. This choice allows the rejection of intrinsic noise while eliminating a negligible fraction of hadronic showers produced by pions of energy larger than $5~\GeV$. The information related to the coordinates of the hits, determined from the location of the fired pad and the plane to which it belongs, is saved together with the threshold reached (either 1, 2 or 3).

\subsection{Data quality control}
\label{sec:DQ}

A dual data quality (DQ) controls was performed, online and offline. The online DQ uses the fast calculation of the efficiency of each chamber and of the occupancy of each channel to isolate the noisy channels and mask them at the very front end electronic level. 

The offline DQ control is focused on the data analysis by measuring the noise, and the performances of each sensitive layer of the detector.

\begin{figure}[!htbp]
  \centering
  \begin{subfigure}[b]{0.35\textwidth}
    \centering
    \includegraphics[width=\textwidth]{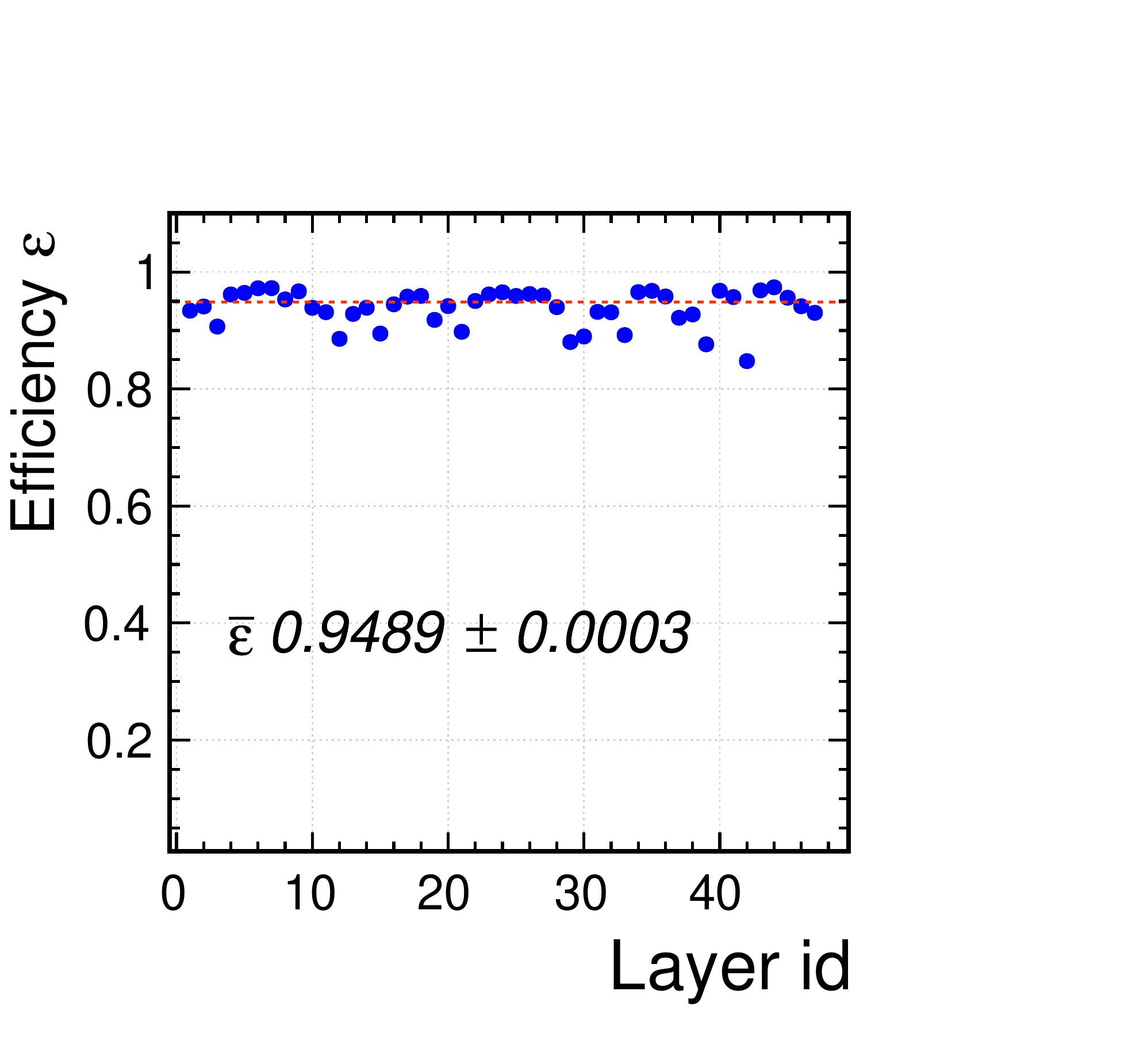}
    \vspace{-1cm}
    \caption{ }
    \label{fig:eff}
  \end{subfigure}%
  \begin{subfigure}[b]{0.35\textwidth}
    \centering
    \includegraphics[width=\textwidth]{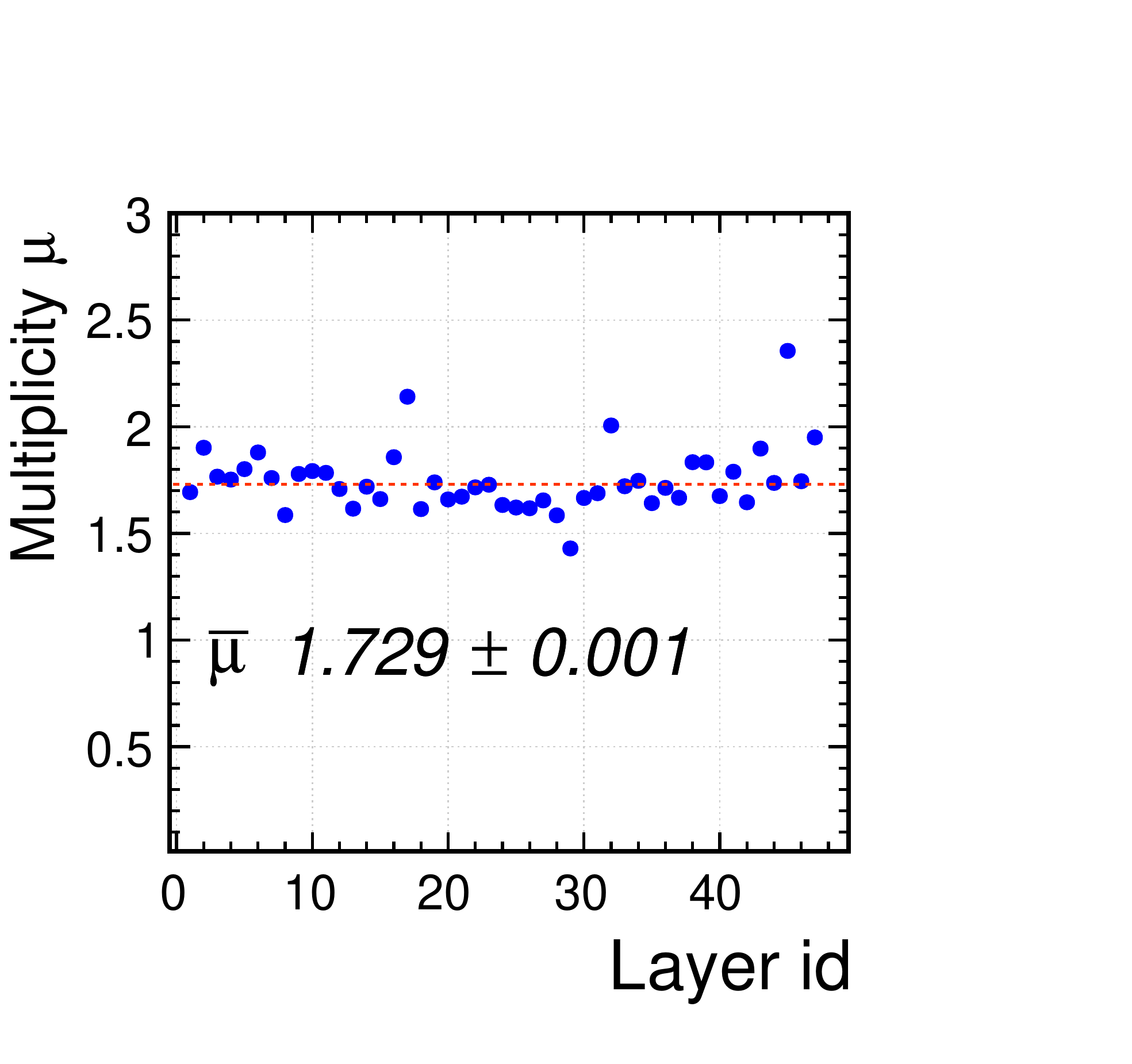}
    \vspace{-1cm}
    \caption{ }
    \label{fig:mul}
  \end{subfigure}%
  \caption{
    Efficiency ( {\bf a} ) and particle multiplicity
    ( {\bf b} ) of the 48 layers.  The red line is the average value.}
  \label{plot:eff_mu}
\end{figure}

The detector performances can be summarized in two quantities: the efficiency and pad multiplicity. They are estimated for each of the 48 layers using the beam muon tracks. Prior to the reconstruction of the muon track, a local clustering was performed in each layer by merging the hits sharing an edge. Isolated clusters are dropped \footnote{ the minimum distance of a cluster must be $ < 12 ~\rm cm $ from the others.}. The tracks are then built using the remaining clusters by minimizing a $\chi^2$. Only tracks with $\chi^2 < 20$ are kept. The efficiency is defined as the probability to find at least 1 hit within $3~\rm cm$ of reconstructed track. The multiplicity is defined as the mean number of hits matched on the considered layer. Figures \ref{fig:eff} and \ref{fig:mul} show respectively the efficiency and the multiplicity of the calorimeter layers.

Two kinds of noises can be distinguished: intrinsic and coherent noises. Intrinsic noise is defined as the hits outside an event (figure \ref{fig:time_spectrum}), due to gain fluctuation in GRPC's.
Its intensity is a function of temperature and the polarization high voltage. The figure \ref{fig:noise_out_event} shows the distribution of the noise in the unit of the time-stamp with an average of $\sim 0.35 ~\rm hit/200ns$. This measurement gives an estimation of the contamination of noisy hits along the physical events and shows that the GRPC are almost noise free. Among the reconstructed events some are clearly due to electronic noise. These events are characterized by the occurrence of many hits belonging to the same electronics slab or to whole electronic layer (made of three slabs). They appear with an incidence of $\sim2/10^6$ events and they are easy to identify thanks to their specific topology.

\section{Particle identification \& Event selection}
\label{sec:particle_id}

As a result of the trigger-less mode (section \ref{sec:trivent}), a large fraction of muon and electron contaminates\footnote{ For beam energies over $20~\GeV$, a lead absorber of $4~mm$ thickness is used to reduce the electron contamination and it is not used for the energies below $20\GeV$ to save the statistic.} the pion samples. Several variables based on topological characteristics of the single particle showers are used to select the hadronic ones and measure their energy. The variables are computed using the information of all hits for a given event.

\begin{figure}[!htbp]
  \centering
  \begin{subfigure}[b]{0.4\textwidth}
    \centering
    \includegraphics[width=\textwidth]{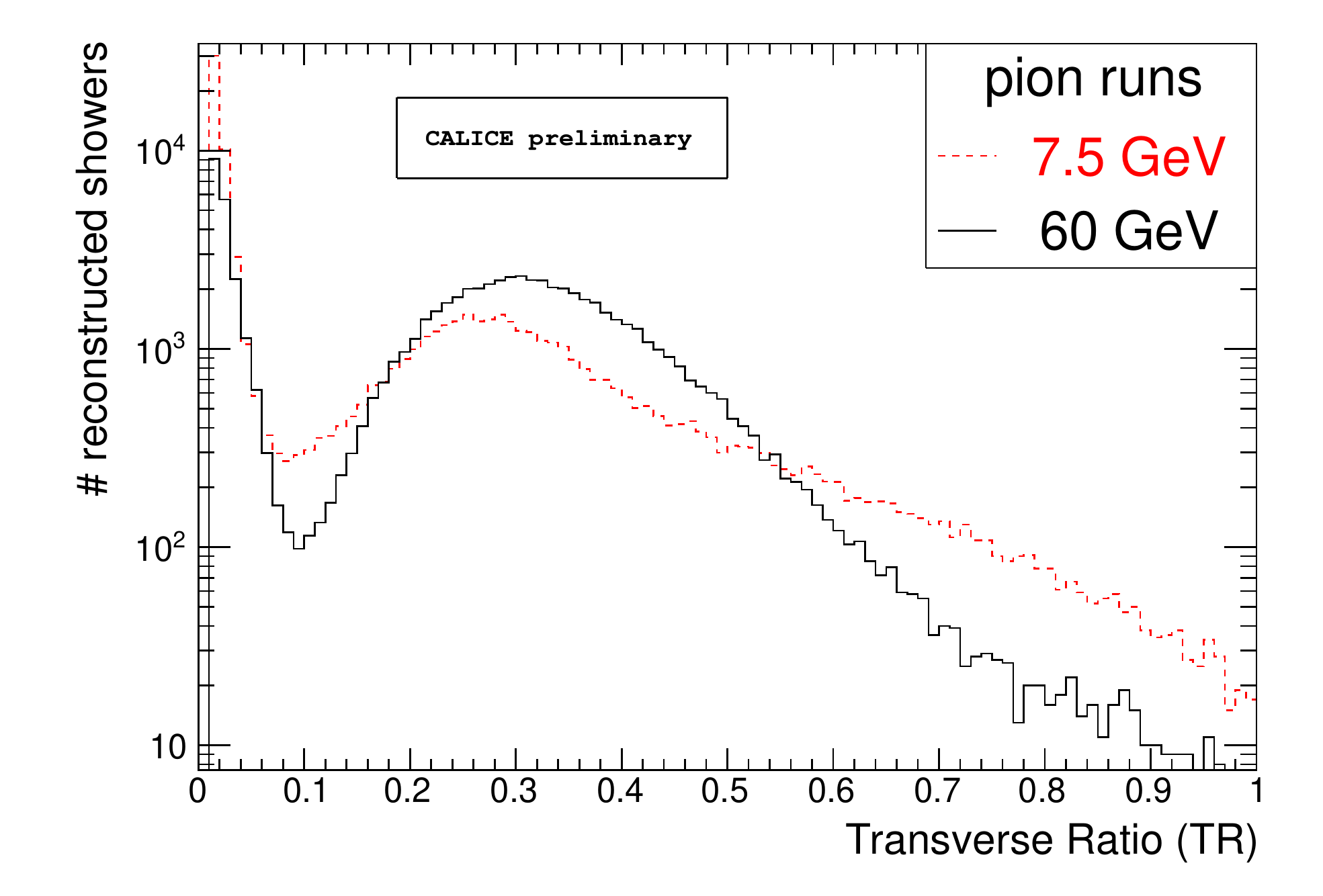}
    \vspace{-1cm}
    \caption{ }
    \label{fig:TR}
  \end{subfigure}%
  \hspace{1cm}
  \begin{subfigure}[b]{0.4\textwidth}
    \centering
    \includegraphics[width=\textwidth]{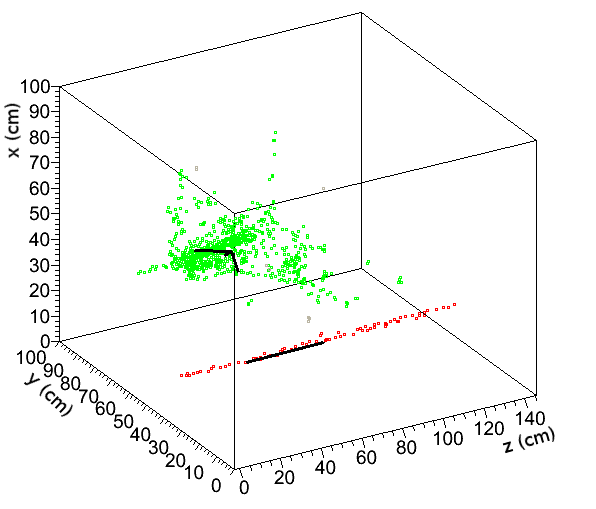}
    \vspace{-1cm}
    \caption{ }
    \label{fig:event_display}
  \end{subfigure}%
  \caption{
    Transverse ratio ({\bf a}) distribution for two different energies. The rejection cut $TR < 0.1$ provides a good muon rejection as shown in ({\bf b}). }
\end{figure}

A Principal Component analysis (PCA) was first applied \cite{method:PCA}: the event's principal axes are calculated and corresponding eigenvalues $\lambda_{i}$ equal to the standard deviation of the hits distribution along the axis i, are sorted in increasing order ($\lambda_{1}<\lambda_{2}<\lambda_{3}$). A transverse ratio ($TR = (\lambda_{1} \oplus \lambda_{2})/\lambda_{3}$) is defined and used to distinguish muons
shape from the interacting particles (figure \ref{fig:TR} and \ref{fig:event_display}). The value of the $TR$ is expected to be close to 0 for the muons. A	 rejection cut at $TR>0.1$ removes more
than $98\%$ of muons.

Additional variables are introduced to reduce the electron contamination. The first variable is defined as $V_1 = (\sum_{layer}N^{layer}_{25})/N_{hit}$, where $N^{layer}_{25}$ is the number of hits in $5\times 5$ pads around the hits barycenter in given layer and $N_{hit}$ represents the total number of hits in the event \footnote{this variable is expected to be greater for electromagnetic showers than for hadronic ones.}. The second variable is defined as $V_2 = D_{f}/\ln(N_{hit})$. The $D_{f}$ is the ``Fractal Dimension" as described in \cite{FractalDim} which characterizes the self-similarity of objects. The determination of this quantity is based on the box-counting algorithm, which consists in counting the number of hits by varying the virtual cell size.%% maybe said more about FD

\begin{figure}[!htbp]
  \centering
  \begin{subfigure}[b]{0.4\textwidth}
    \centering
    \includegraphics[width=\textwidth]{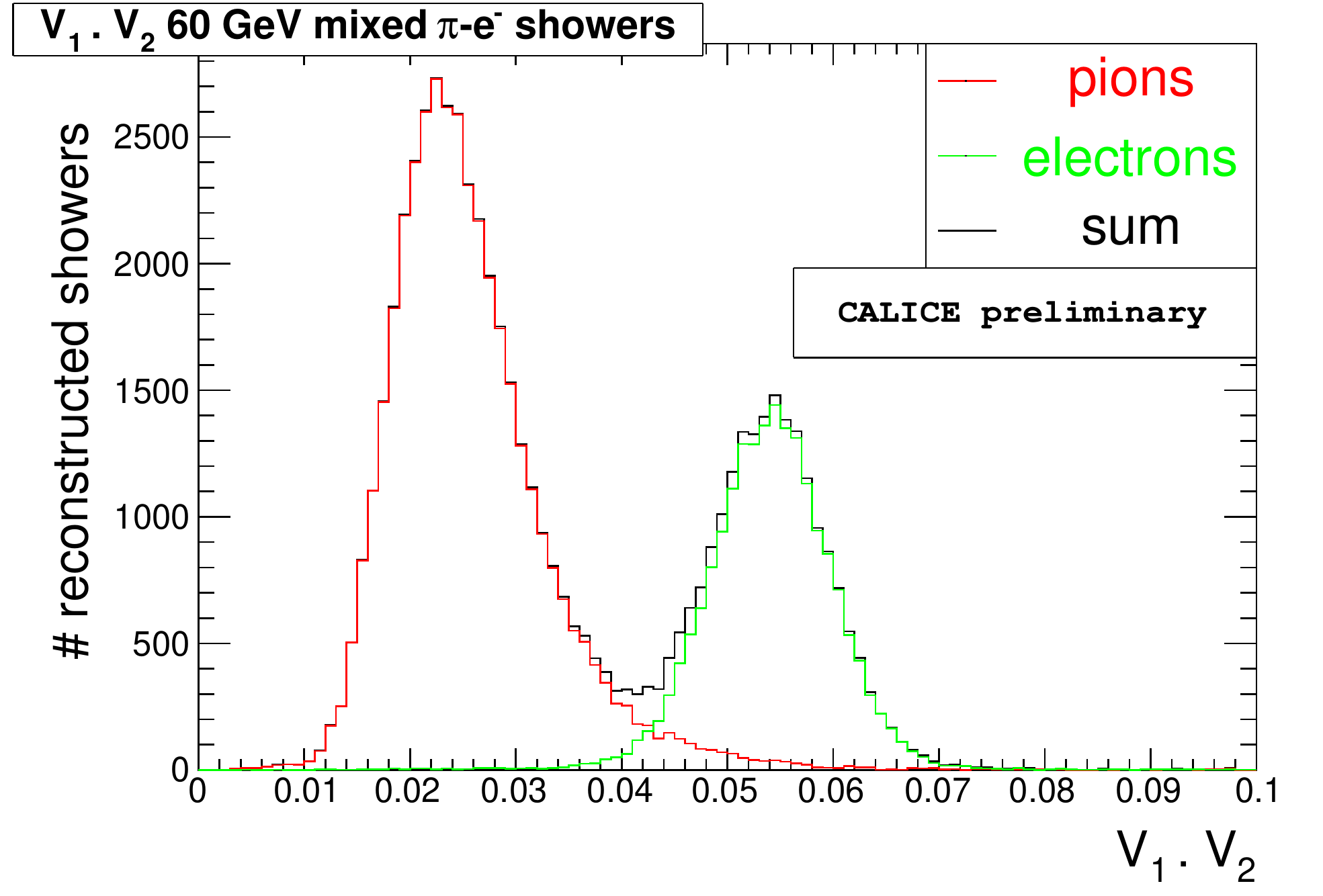}
    \vspace{-1cm}
    \caption{ }
    \label{fig:mix_e_pi_60}
  \end{subfigure}%
  \hspace{1cm}
  \begin{subfigure}[b]{0.4\textwidth}
    \centering
    \includegraphics[width=\textwidth]{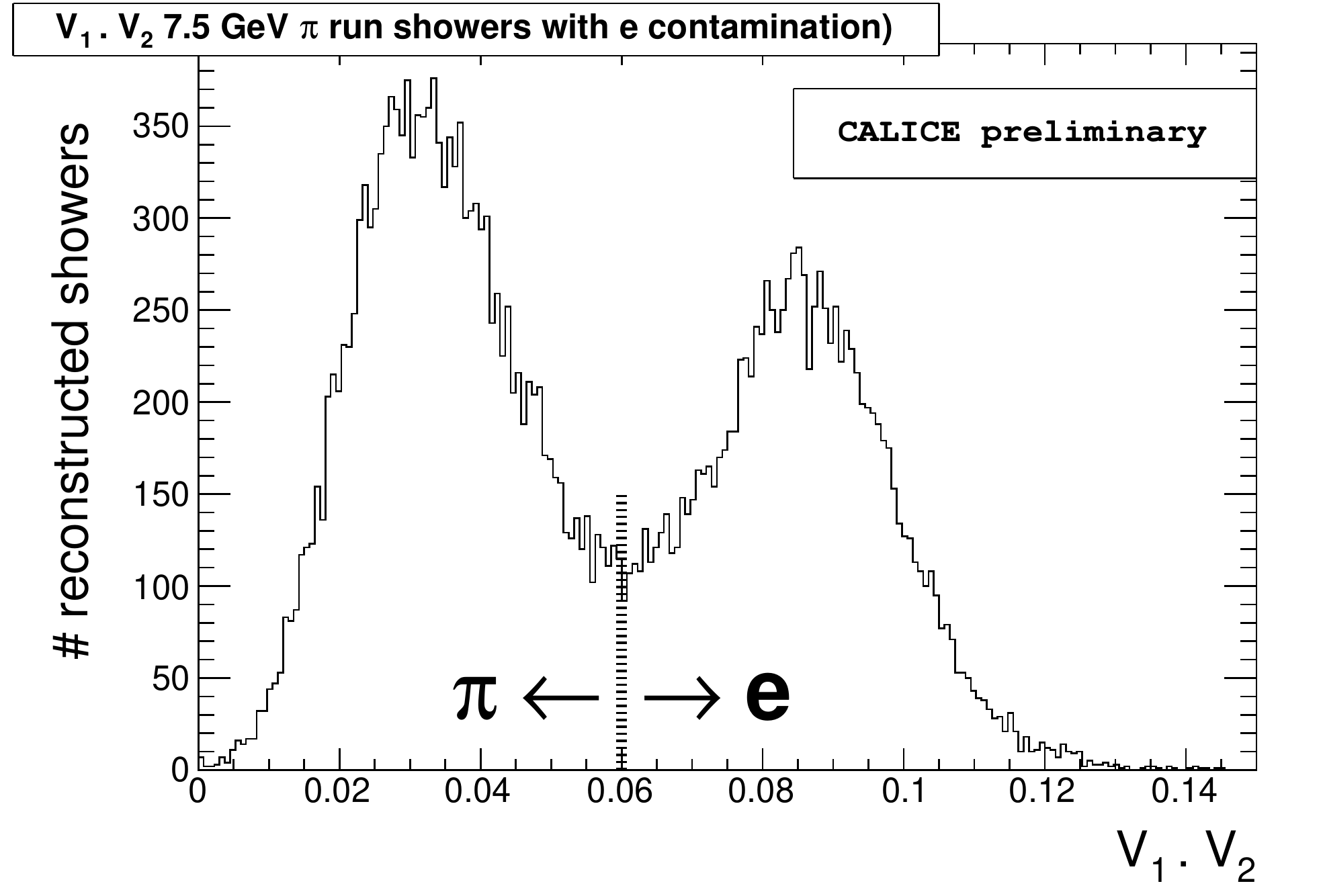}
    \vspace{-1cm}
    \caption{ }
    \label{fig:mix_e_pi_7.5}
  \end{subfigure}%
  \caption{ $V_1 \cdot V_2$ distribution ({\bf{a}}) for pion (red) and electron (green) runs with same energy ($60\GeV$). ({\bf{b}}) A cut about $V_1 \cdot V_2 = 0.05$ at $7.5\GeV$ provides $e/\pi$ separation.}
  \label{fig:mix_e_pi}
\end{figure}

% FIXME: a reformuler 
To reach a better electromagnetic and hadronic shower separation, the two topological variables introduced previously are combined. As seen in figure \ref{fig:mix_e_pi_60}, the product of the variables $V_1\cdot V_2$ allows for a clear separation between electron and pion samples of the same energy between $7.5 $ and $60~\GeV$. The table \ref{tab:v1v2_cut} summarizes the cut applied at different energies.

\begin{table}[h!]  
  \begin{center}
    \begin{tabular}{|c|c|c|c|c|c|c|} 
      \hline
      $E_{beam}$ & 5 & 7.5 - 15 & 20 - 25 & 30 - 40 & 50 - 60 & 70 - 80 \\ 
      \hline
      cut ($V_1 \cdot V_2$) & 0.065 & 0.06 & 0.55 & 0.05 &0.045 & 0.04 \\
      \hline
    \end{tabular} 
    \caption{Summary of the optimal cut on $V_1 \cdot V_2$ at different energies, to separate electrons from pions.} 
    \label{tab:v1v2_cut} 
  \end{center}
\end{table}

Finally the following constraints are applied to select well
contained showers:
\begin{itemize} %% ICI
\item The first interacting plate \footnote{The first interacting layer is defined as the first layer having more than 5 hits followed by at least 3 plates having also more than 5 hits.} must be in the first 15 layers to exclude the late interacting hadrons.
\item The ratio of the number of hits in the last 7 layers over the one in first 30 layers must be less than $15~\%$.
\end{itemize}

\section{Energy Response}\label{sec:energy}
The selected hadronic showers as described in the previous section\ref{sec:particle_id} are used to study the linearity and the energy resolution in the pure and semi-digital modes.	
\subsection{Binary mode}\label{sec:binary}	
	
\begin{figure}[!htbp]
  \centering
  \hspace{-0.8cm}
  \begin{subfigure}[b]{0.35\textwidth}
    \centering
    \includegraphics[width=\textwidth]{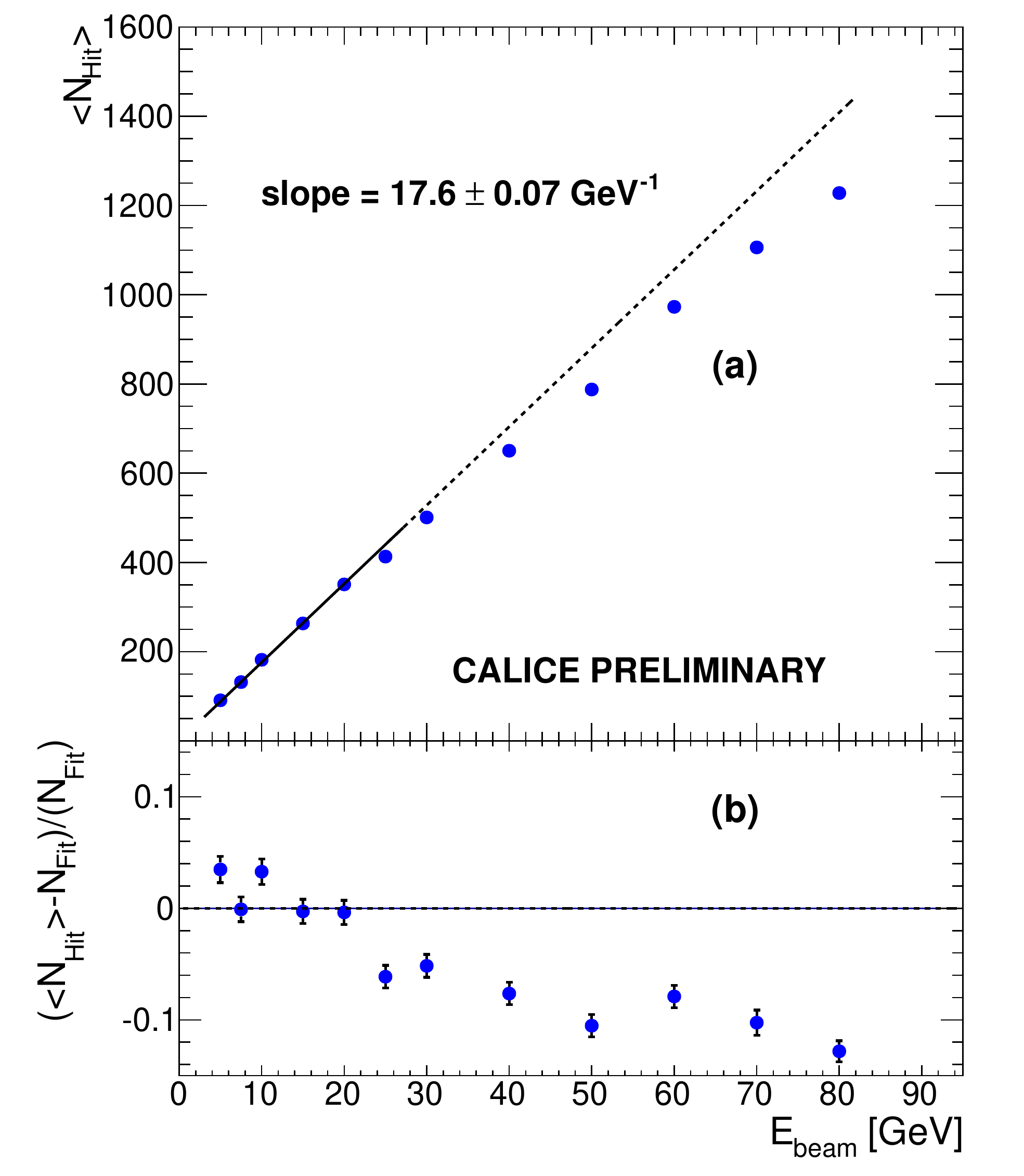}
    \vspace{-1cm}
    \caption{ }
    \label{fig:lin_nhit}
  \end{subfigure}%
  \begin{subfigure}[b]{0.35\textwidth}
    \centering
    \includegraphics[width=\textwidth]{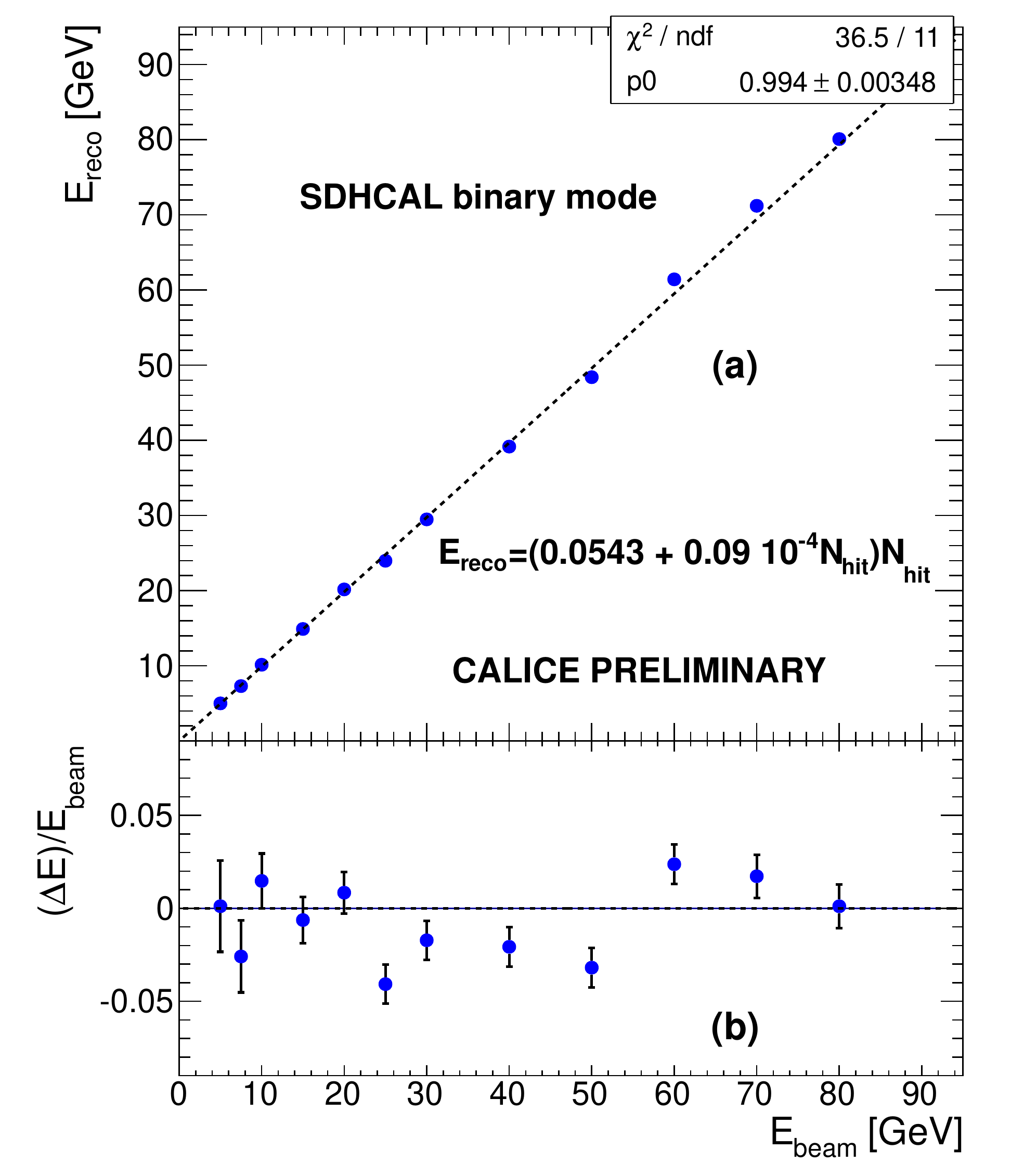}
    \vspace{-1cm}
    \caption{ }
    \label{fig:lin_dhcal}
  \end{subfigure}%
  \begin{subfigure}[b]{0.35\textwidth}
    \centering
    \includegraphics[width=\textwidth]{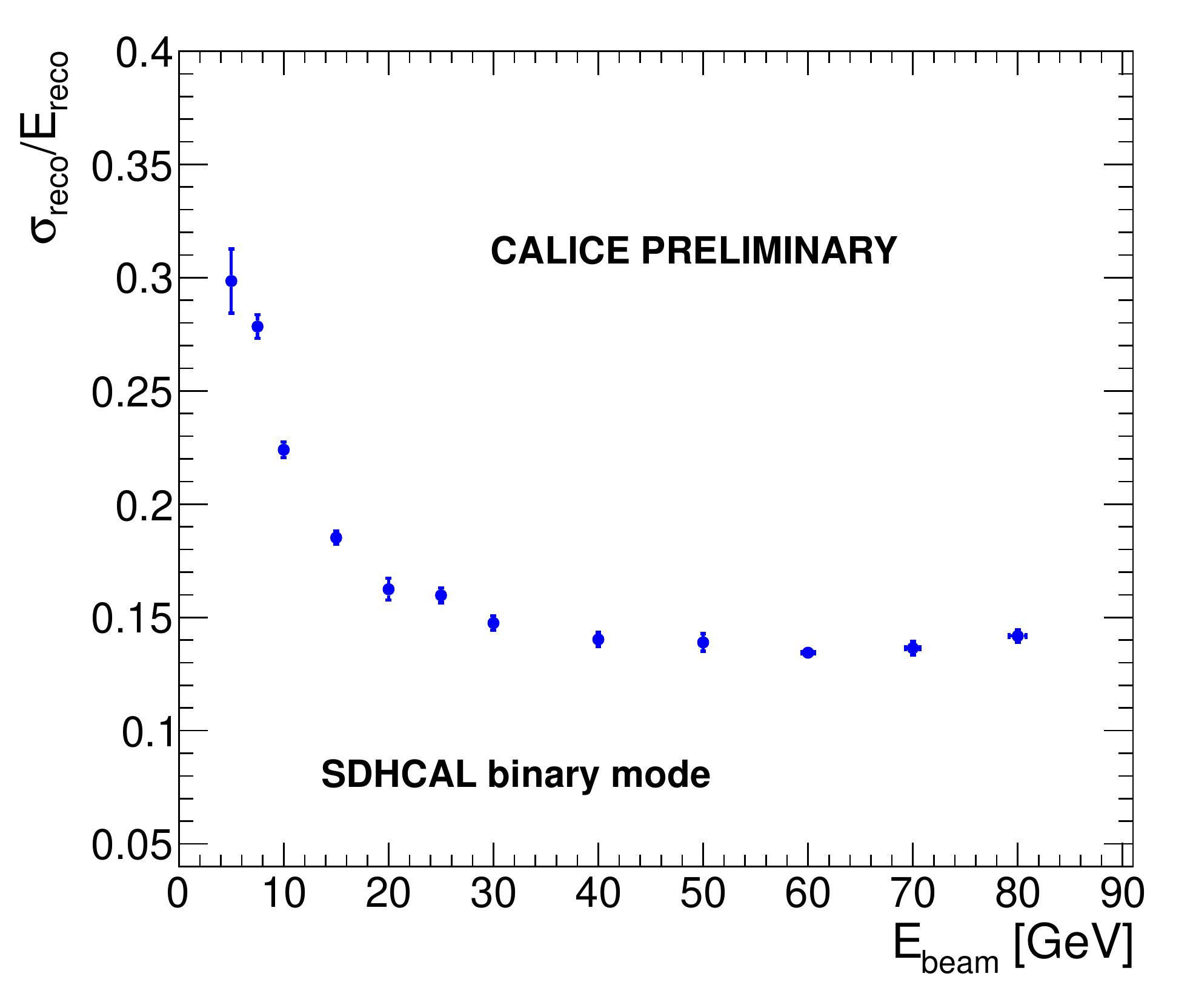}
    \vspace{-1cm}
    \caption{ }
    \label{fig:res_dhcal}
  \end{subfigure}%
  \label{fig:lin}
  \caption{(a) A fit is applied between 5 and $25 \GeV$ using a linear
    parametrization $E_{\rm reco}=C\cdot N_{hit}$. (b) , (c) Evolution of
    resolution in function of the beam energy}
\end{figure}

As mentioned in the introduction\ref{sec:intro}, the reconstructed energy of hadrons in the pure digital mode is proportional to mean number of hits at first order. The distribution of this quantity ($N_{\rm hit}$) is fitted for each beam energy using a Crystal-ball function to extract the average number of hits and the spread \footnote{The choice of the Crystal-ball function can be justified to take into account the tail at the low number of hits.}. The figure \ref{fig:lin_nhit} shows the average number of hits versus beam energy. The non-linearity, observed for the $E_{\rm{beam}}\geq 30~\GeV$, is due essentially to the saturation effect at high energy. In order to improve the estimation of the reconstructed energy, a quadratic function of $N_{\rm nhit}$ is then chosen: $E_{\rm reco} = (C+D\cdot N_{\rm hit}) N_{\rm nhit}$. The parameters $C$ and $D$ are derived from a minimization of $\chi^2=\sum^{N_{\rm event}}_{i} (E_{\rm beam}^i - E_{\rm reco}^i)^2/E_{\rm beam}^i$.  

The energy resolution is defined as $\sigma(E_{\rm reco})/E_{\rm reco}$ where the $\sigma(E_{\rm reco})$ is extracted from crystal ball fit to the reconstructed energy distribution; it is shown on figure \ref{fig:res_dhcal} as function of the beam energy.  

The quadratic model of the energy reconstruction, restores the linearity up to $80~\GeV$ (figure \ref{fig:lin_dhcal}). But the saturation effect still impact the tail of the energy resolution at energies over 50 GeV (figure \ref{fig:res_dhcal}).
 
\subsection{Semi-Digital mode}\label{sec:semidigital}
\begin{figure}[!htbp]
  \centering
  \begin{subfigure}[b]{0.35\textwidth}
    \centering
    \includegraphics[width=\textwidth]{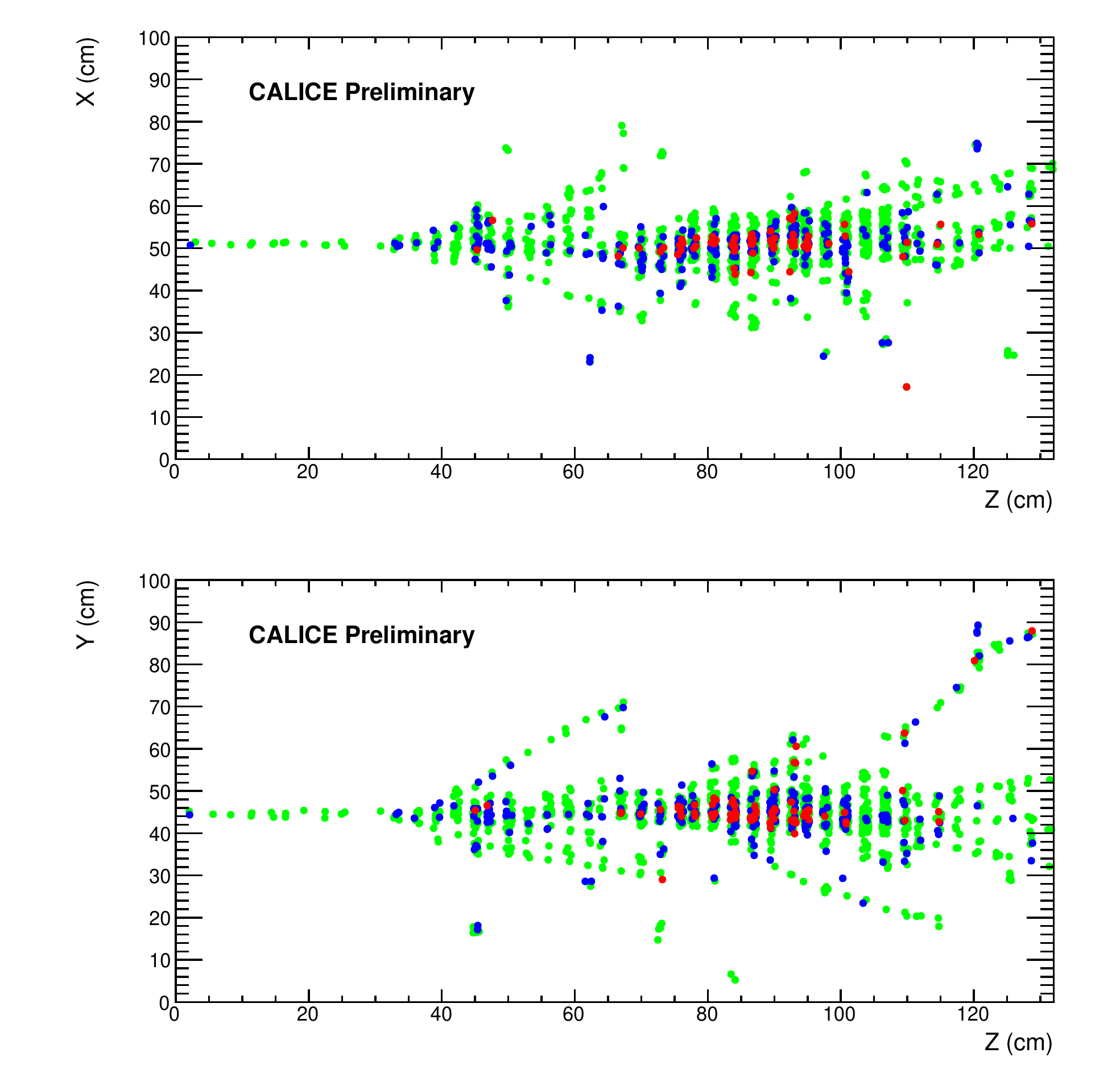}
    \vspace{-1cm}
    \caption{ }
    \label{fig:evt_pi80}
  \end{subfigure}%
  \begin{subfigure}[b]{0.35\textwidth}
    \centering
    \includegraphics[width=\textwidth]{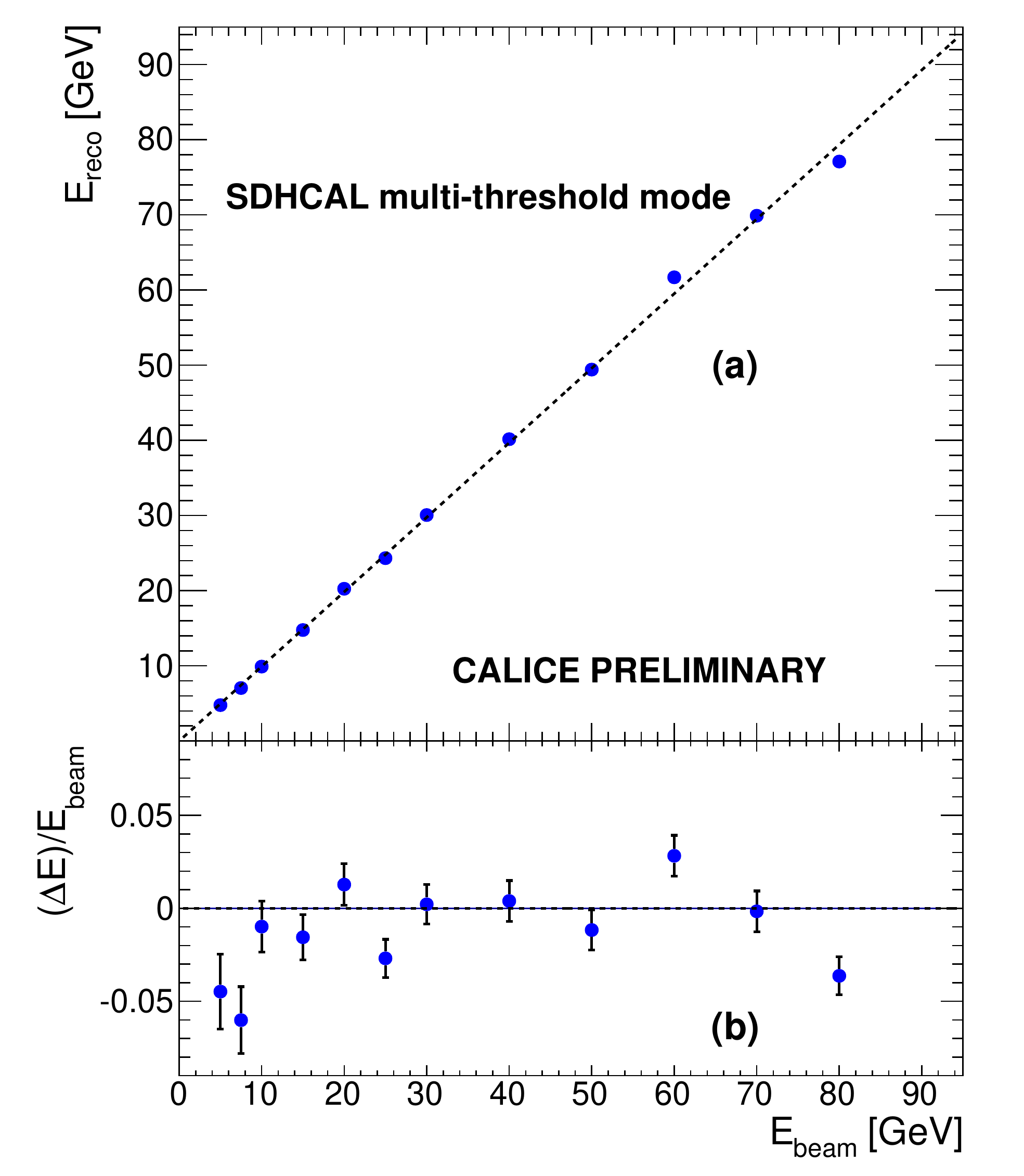}
    \vspace{-1cm}
    \caption{ }
    \label{fig:lin_sdhcal}
  \end{subfigure}%
  \begin{subfigure}[b]{0.35\textwidth}
    \centering
    \includegraphics[width=\textwidth]{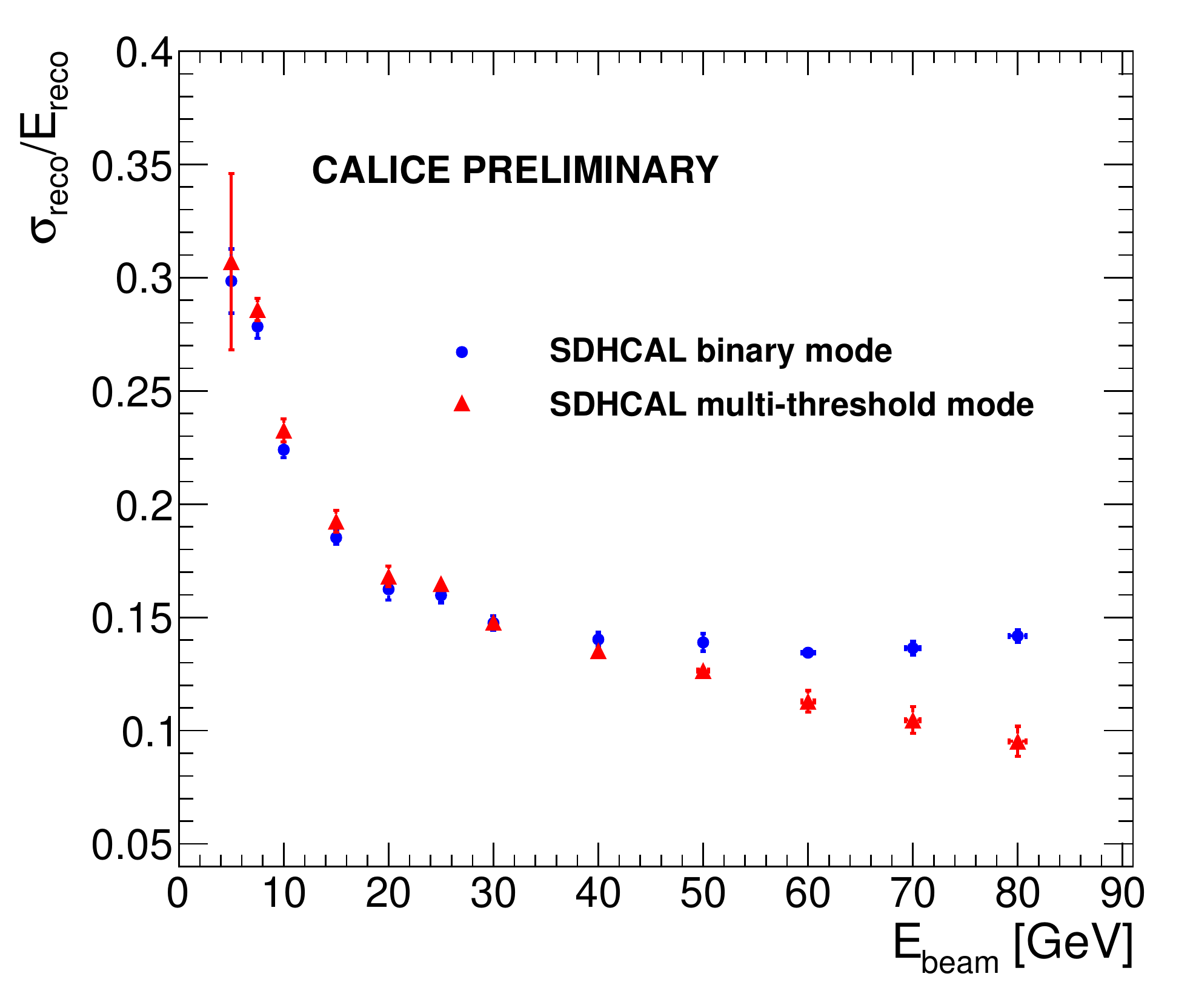}
    \vspace{-1cm}
    \caption{ }
    \label{fig:res_sdhcal}
  \end{subfigure}%
  \caption{ ({\bf a}) event display(top and side) of pion shower inside the calorimeter. The colors (green, blue, red) correspond to the different thresholds (1,2,3). ({\bf b}) Reconstructed energy versus beam energy in semi-digital mode. ({\bf c}) Energy resolution for pions in the binary (blue) and semi-digital mode (red).}
  % \label{fig:sdhcal_mode}
\end{figure}

% refaire ce text
The semi-digital case as described previously is characterized by the presence of three thresholds corresponding to three levels of deposit energy. This configuration leads to a better estimation of the energy when the showers are very dense (figure \ref{fig:evt_pi80}). Indeed, in the shower core the density of the particle is very high and the saturation can be reached easily in the digital case. However, the presence of the three thresholds can provide additional information which may overcome this issue. 

The reconstructed energy can be defined as a simple linear combination of number of hits at each threshold. $E_{\rm reco} = \alpha N_1 + \beta N_2 + \gamma N_3$, the $N_i$ are the number of hits for a given threshold \footnote{ The number of hits for each threshold is exclusive: $N_1$ is the number of hit crossing only the first threshold, $N_2$ is the number of hits crossing the second threshold at the exclusion of the first and third one. $N_3$ is the number of hits above the third threshold.}. The parameters $\alpha$, $\beta$ and $\gamma$ are parametrised as quadratic functions of the total number of hits, determined using the same method as described in the previous section. The linearity as shown in figure \ref{fig:lin_sdhcal} is respected for the full range with an accuracy of $5\%$. 

The energy resolution as represented in the figure \ref{fig:res_sdhcal} for the semi-digital case improves the energy resolution by $30\%$ at $80~\GeV$.

\section*{Conclusion}

The CALICE technological SDHCAL prototype using auto-trigger and power-pulsed very frond end embedded electronics, was tested during the 2012 beam test runs. The data quality was verified online and offline. The data incorporating all contributions (beam pions, muons, electrons and cosmics) were cleaned up and pions were clearly selected using a data driven set of cuts. The resolution associated to the linearised energy response to pion of $5$ to $80~\GeV$, was estimated in both binary (lowest threshold used) and semi-digital (tree thresholds used) modes. The multi-threshold mode of the SDHCAL improves the resolution at high energy (>$50~\GeV$)  by up to $30~\%$ at $80~\GeV$, thanks to better treatment of saturation effect using the information provided by the second and the third thresholds (4 and 15 mip respectively).

\bibliography{mybib}{}
\bibliographystyle{utphys}

\end{document}